\documentclass[12pt,a4paper]{article}

\usepackage{amssymb,amsmath,eucal}
\usepackage[dvips]{lscape,graphicx}

\newcommand{\bc}{\begin{center}}
\newcommand{\ec}{\end{center}}
\newcommand{\bd}{\begin{displaymath}}
\newcommand{\ed}{\end{displaymath}}
\newcommand{\be}{\begin{equation}}
\newcommand{\ee}{\end{equation}}
\newcommand{\ba}{\begin{array}}
\newcommand{\ea}{\end{array}}
\newcommand{\bt}{\begin{tabular}}
\newcommand{\et}{\end{tabular}}

\newcommand{\ds}{\displaystyle}

\newcommand{\lb}{\label}
\newcommand{\bp}{\begin{picture}}
\newcommand{\ep}{\end{picture}}
\newcommand{\bfi}{\begin{figure}}
\newcommand{\efi}{\end{figure}}
\begin{document}
\title{\bf Semi-analytical solution for Bogolubov's angle $\theta(p)$ in the 't~Hooft model}

\date{\it School of Physics and Astronomy, University of Minnesota, \\
Minneapolis, MN 55455, USA}
\author{Dmitry Ryzhikh\thanks{ryzhikh@physics.umn.edu}}

\maketitle

\begin{abstract}
The analytical ansatz was found for Bogolubov's angle $\theta(p)$
in the 't~Hooft model at $N\rightarrow\infty$. The appropriate
calculations were done and final numeric approximation was found
for this angle $\theta(p)$.
\end{abstract}
\thispagestyle{empty}%


%
\newpage
\section{Introduction}
There is a well known generalization of QCD from three colors to
$N$ ($N\in\mathbb{N}$) colors or, in other words, from an $SU(3)$
gauge group to an $SU(N)$ gauge group which was done by
t'Hooft~\cite{tHt}. t' Hooft made this generalization because he
hoped that model with a large $N$ (strictly speaking with limit
$N\rightarrow \infty$, the so-called t'Hooft limit) might have an
exact solution and might be qualitatively and quantitatively close
to the actual QCD with $N=3$. His hope was justified; theory in
the $N\rightarrow \infty$ limit has a considerable simplification.
The fact is that we should take into account only planar diagrams,
because others vanish~\cite{mas}: {\it ``In the actual world
quarks belong to the fundamental representation of $SU(3)$. If we
assume that this assignment stays intact in multicolor QCD, each
extra quark loop is suppressed by $1/N$. Therefore, in the
't~Hooft limit each process is saturated by contributions with the
minimal possible number of quark loops"}. This is the main reason
we are motivated to investigate this model and to find a solution
for Bogolubov's angle. %

All definitions and formulas which were necessary for presented
calculations were taken from the paper~\cite{mas}.

\section{Analytical ansatz for $\theta(p)$}

The starter formulas for further calculations:

\be%
\lb{eqs}%
\ba{l}%
E_p\cos\theta(p)-m=
\dfrac{\gamma}{2}\int dk\cos\theta(k)\dfrac{1}{(p-k)^2}\, ,\\
E_p\sin\theta(p)-p= %
\dfrac{\gamma}{2}\int dk\sin\theta(k)\dfrac{1}{(p-k)^2}\, ,
\ea%
\ee%
with the boundary conditions
\be%
\lb{boundcond}%
\theta(p)\rightarrow \left\{
\ba{l}%
\phantom{-}\dfrac{\pi}{2} \quad\mbox{at}\quad
p\rightarrow\infty\\[0.3cm]
-\dfrac{\pi}{2}\quad\mbox{at}\quad p\rightarrow-\infty
\ea%
\right.
\ee%
determined by the free-quark limit. This set of equations was
firstly obtained by Bars and Green~\cite{bar}. The angle
$\theta(p)$ is referred to as the Bogolubov's angle, or more
commonly, the chiral angle.

From these equations one can easily get the following integral
equation for the Bogolubov's angle~\cite{bar, len}:
\be%
\lb{theta}%
p\cos\theta(p)-m\sin\theta(p)=\dfrac{\gamma}{2}\int
dk\sin\left(\theta(p)-\theta(k)\right)\dfrac{1}{(p-k)^2}\; .
\ee%
Assuming that the chiral angle was found the following integral
equation for the $E_p$ was obtained in paper~\cite{bar, len}:
\be%
\lb{Ep}%
E_p=m\cos\theta(p)+p\sin\theta(p)+\dfrac{\gamma}{2}\int
dk\cos\left(\theta(p)-\theta(k)\right)\dfrac{1}{(p-k)^2}\; .
\ee%

All calculations were performed in the limit of quark mass $m=0$.
Thus, one can rewrite~Eq.~(\ref{theta}) and~Eq.~(\ref{Ep}) in this
limit.
\be%
\lb{EqT}
p\cos{\theta(p)}=%
\frac{\ds\gamma}{\ds
2}\int\limits_{-\infty}^\infty\frac{\ds{\sin[\theta(p)-\theta(k)]}}{\ds{(p-k)^2}}\,\,dk\; ,%
\ee %
\be%
\lb{Epm0}%
E_p=p\sin\theta(p)+\dfrac{\gamma}{2}\int
\dfrac{\cos\left(\theta(p)-\theta(k)\right)}{(p-k)^2}\,\, dk\; .
\ee%

The exact singular analytical solution of the integral equation~(\ref{EqT}) is %
\be%
\lb{ansol}%
\theta(p)=\frac{\ds\pi}{\ds 2}\,\mbox{sign}(p)\; ,
\ee%
where sign$(p)$ is the sign function %

$$
\mbox{sign}(p)=\vartheta(p)-\vartheta(-p)\; .
$$

\bfi[h]
\bc%
\includegraphics[width=90mm, keepaspectratio=true]{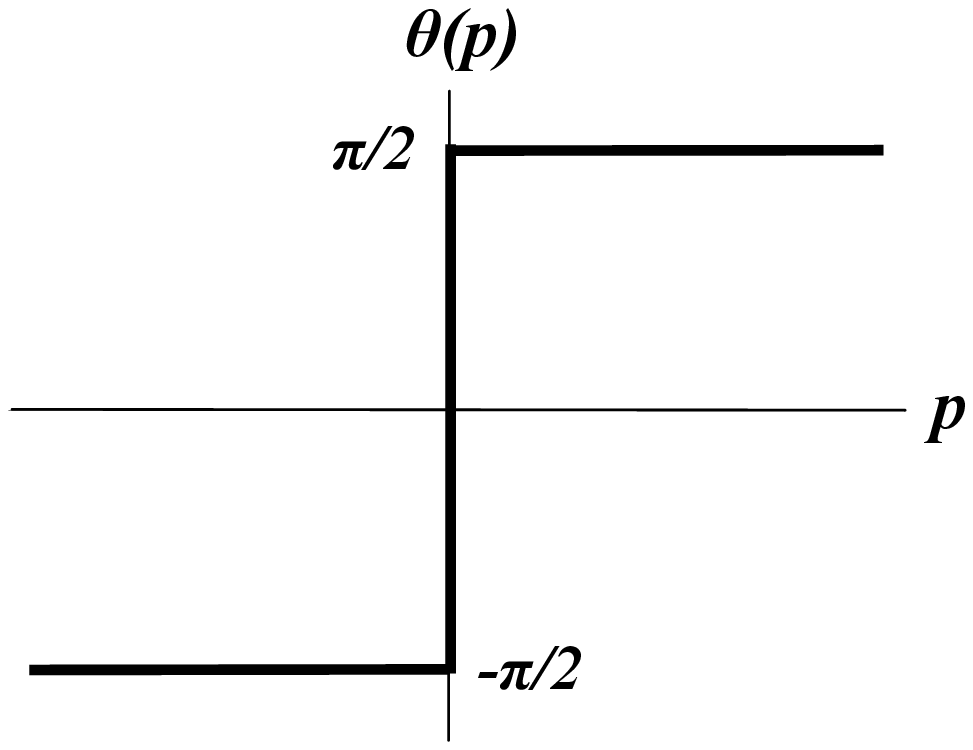}
\ec%
\vspace*{-0.8cm}
\caption{The analytical solution for Bogolubov's angle $\theta(p)$ vs. $p$.}%
\efi

Substituting this solution in~Eq.~(\ref{Epm0}) we can get
\be%
\lb{Epsol}%
E_p=|p|-\dfrac{\gamma}{|p|}\; .
\ee%

However, this analytical solution is unphysical~\cite{mas} for
several reasons. For example, $E(p)$ becomes negative at
$|p|<\sqrt{\gamma}$. This feature of the solution~(\ref{Epsol})
cannot be amended by a change in the infrared regularization. In
fact,~solution~(\ref{Epsol}) does not correspond to the minimum of
the vacuum energy~\cite{kal}.

A stable solution has the form depicted in Fig.~2. It is smooth
everywhere. At $|p|\ll\sqrt\gamma$ it is linear in $p$.
\bfi[h]
\bc%
\includegraphics[width=100mm, keepaspectratio=true]{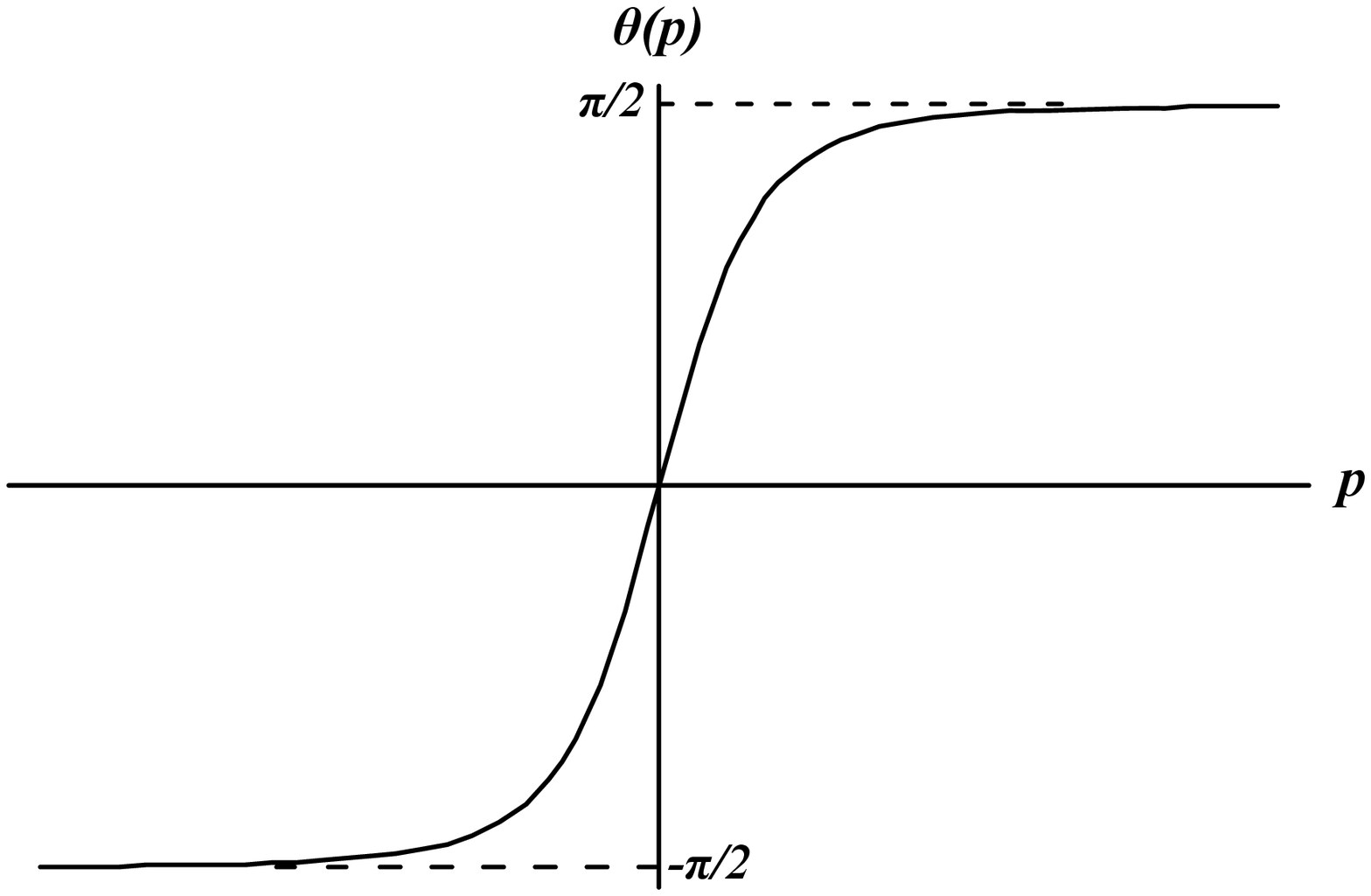}
\ec%
\vspace*{-0.8cm}
\caption{A stable solution for Bogolubov's angle $\theta(p)$ vs. $p$.}%
\efi

For the convenience of the further calculations, the system of
units in which $\gamma=1$ was used. The asymptotic behavior of the
physical $\theta(p)$~\cite{mas}:
\be%
\lb{asbeh1}%
\ba{lll} %
\theta(p)\sim p&\mbox{for } |p| \ll1 &\\[-0.25cm]
&&\mbox{and}\\[-0.25cm]%
\theta(p)= \frac{\ds\pi}{\ds 2}\,\mbox{sign}(p)-\frac{\ds\pi}{\ds\sqrt{6}}\left(\frac{\ds1}{\ds p}\right)^3+\cdots%
&\mbox{for } |p| \gg1\; .&%
\ea%
\ee%
The asymptotic behavior of $\theta(p)$ at $|p|\gg1$ was calculated
by calculating the chiral quark condensate
$\langle\bar\psi\psi\rangle$ for the smooth physical solution as a
self-consistency condition~\cite{zhi}.

Taking these asymptotics into account, we can find the analytical
function for $\theta(p)$ which we can use to fit physical
$\theta(p)$. Therefore the analytical ansatz for $\theta(p)$ is
\be%
\lb{anz}%
\boxed{%
\;\theta(p)=\frac{\ds 6ap}{\ds 2+\pi\sqrt{6}a^3+6a^2p^2}%
+\arctan{(ap)}}\; , %
\ee%
where $a$ is a free parameter and $a>0$. \\%
For any $a$, this ansatz gives the exact asymptotic behavior of
the physical $\theta(p)$:
\be%
\lb{asbeha}
\ba{rrr} %
\hspace*{-4mm}\!\!\!\!\theta(p)=&&\\
&\hspace*{-14.6mm}=\left(1+\dfrac{6}{2+\pi\sqrt{6}a^3}\right)ap+%
\dfrac{1}{3}\left(1+\dfrac{108}{\left(2+\pi\sqrt{6}a^3\right)^2}\right)a^3p^3+\cdots%
&\mbox{for }|p| \ll1 \;\, \\[-0.15cm]%
&&\mbox{and}\\[-0.6cm]\\%
\hspace*{-4mm}\theta(p)=&&\\
&\hspace*{-16mm}=\frac{\ds\pi}{\ds2}\,\mbox{sign}(p)-%
\frac{\ds\pi}{\ds\sqrt{6}}\frac{\ds1}{\ds{p^3}}+%
\dfrac{5\pi a^3\left(3\pi a^3+2\sqrt{6}\right)-8}{90}\cdot%
\frac{\ds1}{\ds{a^5 p^5}}+\cdots%
&\mbox{for }|p| \gg1\,.%
\ea%
\ee%
The graph of the physical $\theta(p)$ from this ansatz looks
exactly like the curve on~Fig.~2.

\section{Numeric solution of $\theta(p)$}
Mathematica 5.1 was employed to create a computer program for
numerical calculations. Also for these calculations the analytical
ansatz was used and it was found that, within precision of the
calculations, the best fit of the analytical ansatz for
$\theta(p)$ is reached when $a=0.3701$.

The precision of the calculations is $0.0001$. The discrepancy of
a solution of the integral equation~(\ref{EqT}) at $\gamma=1$ is
calculated by the formula:

\be%
\lb{dis}%
\ba{ll} %
\hspace*{-0.5cm}
S(a,p)=&\\
&\hspace*{-1.7cm}=\dfrac{1}{\sqrt{n-1}}\sqrt{\sum\limits^n_{i=0}\left(p_i\cos{\theta_{\mbox{\scriptsize{calc}}}(p_i,a)}-%
\frac{\ds1}{\ds
2}\int\limits_{-\infty}^\infty\frac{\ds{\sin[\theta_{\mbox{\scriptsize{calc}}}(p_i,a)-\theta_{\mbox{\scriptsize{calc}}}(k,a)]}}{\ds{(p_i-k)^2}}\,\,
dk\right)^2}\\ %
\ea%
\ee%
where $n$ is a number of points in which
$\theta_{\mbox{\scriptsize{calc}}}(p_i)$ was calculated.

The dependence of $S(a,p)$ on momentum $p$ is an indirect
dependence. This dependence is realized by the dependence of
$S(a,p)$ on the choice of range of momentum $p$ and on the
fragmentation of this range.

\bfi[h]
\bc%
\includegraphics[width=145mm, keepaspectratio=true]{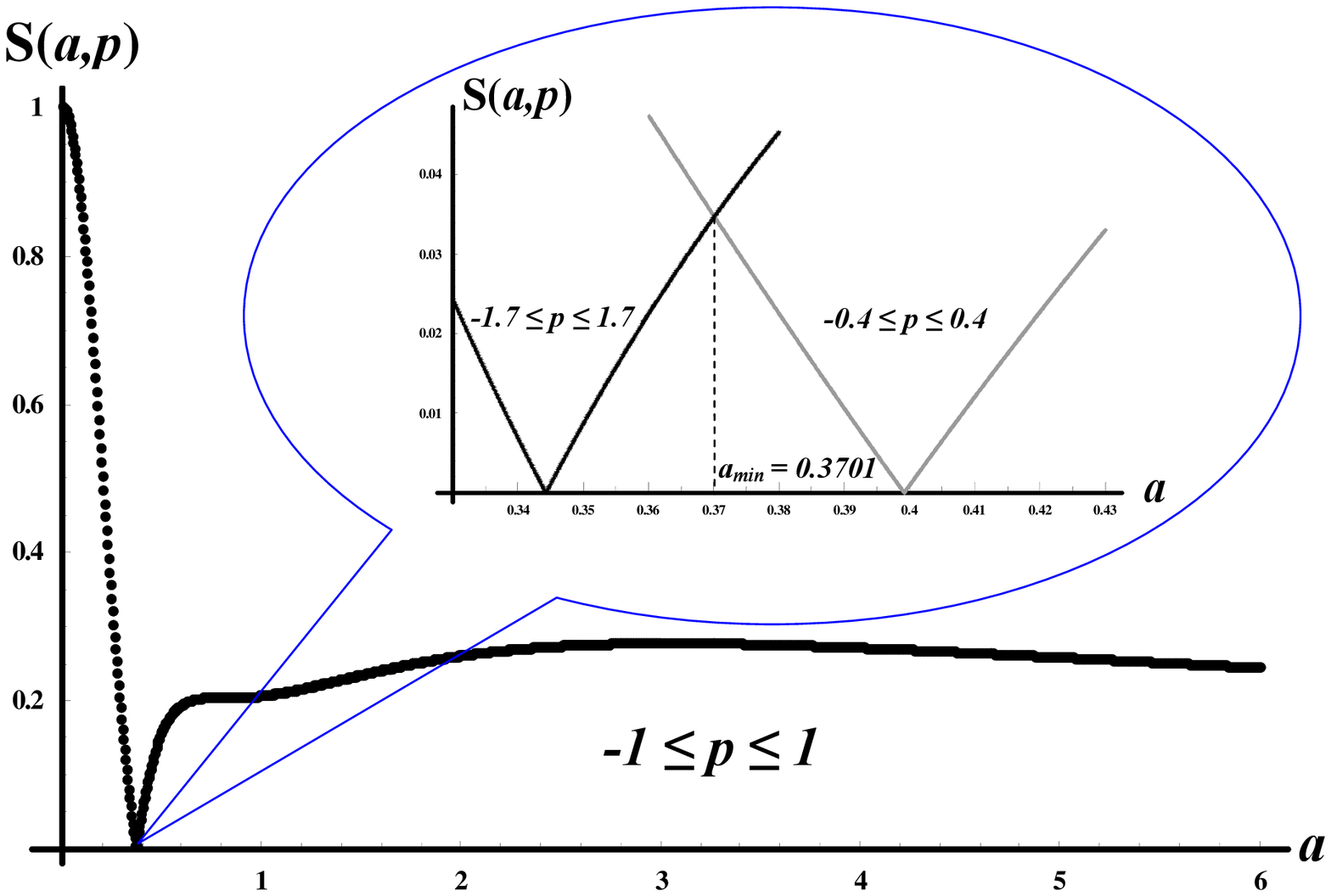}
\ec%
\vspace*{-0.8cm}
\caption{The dependence of the solution discrepancy of the integral equation for $\theta(p)$ on the parameter $a$.}%
\efi%

The calculations of the solution discrepancy  $S(a,p)$ shows that
there are two local maximums of this discrepancy dependent on
momentum $p$ near $p=0.4$ and $p=1.7$ momentum units.
Investigating the dependence of the discrepancy on parameter $a$
at this region of momentum $p$, it was found that the discrepancy
$S(a,p)$ of the calculations is bounded above $0.035$, and that it
reaches the minimum at $a=0.3701$.

For the case $a=0.3701$, the solution for the physical Bogolubov's
angle $\theta(p)$ was obtained:

\be%
\lb{solTh}%
\boxed{
\;\theta(p)=\dfrac{2.2206}{2.3901+0.8218\, p^2}\, p+%
\arctan\left(0.3701p\right)}\;. %
\ee
\bfi[h]%
\bc%
\includegraphics[width=100mm, keepaspectratio=true]{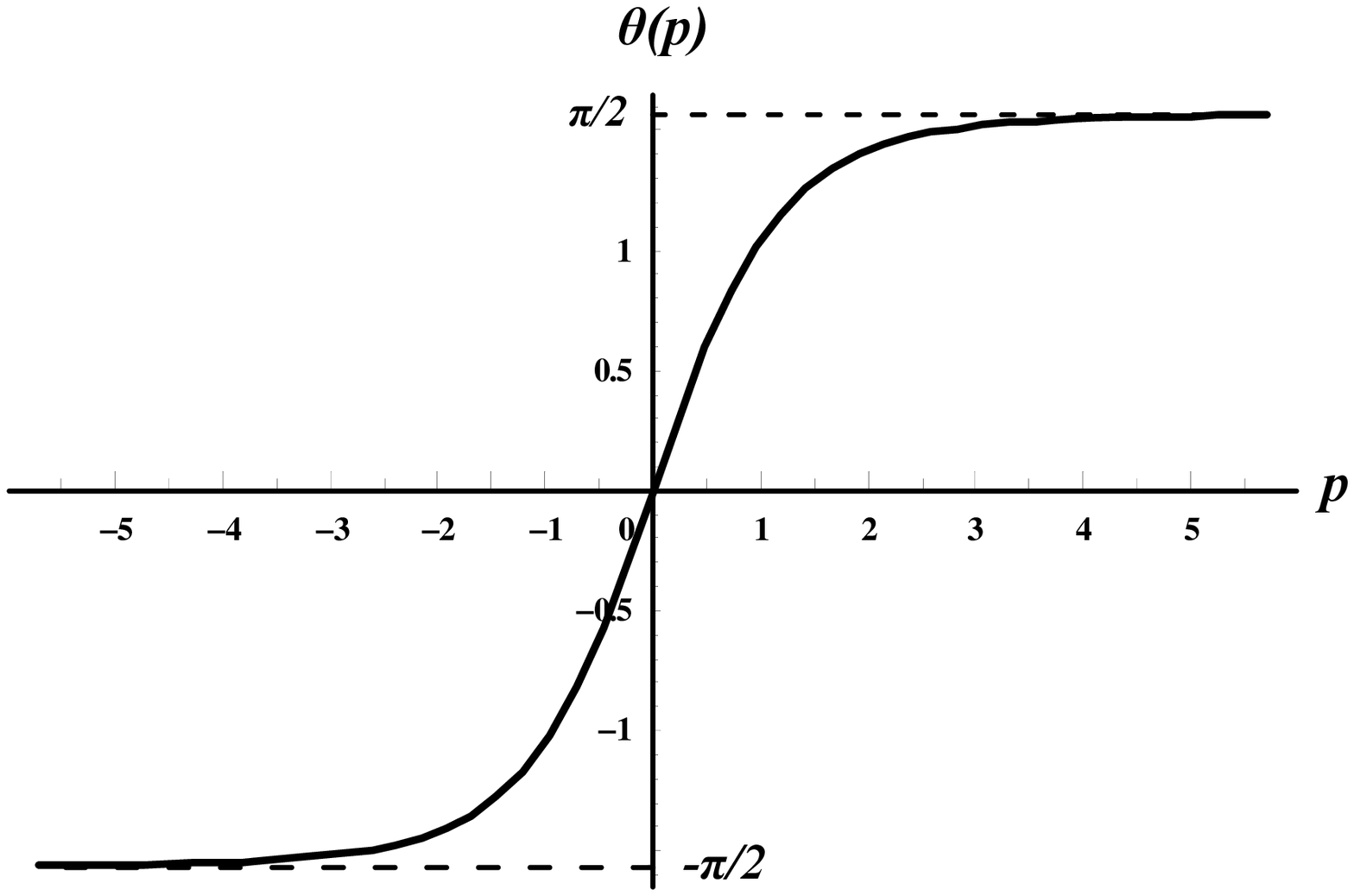}
\ec%
\vspace*{-0.5cm}
\caption{The best fit at $a=0.3701$ for the physical solution of the Bogolubov's angle $\theta(p)$ vs. $p$.}%
\efi

The asymptotic behavior of physical $\theta(p)$:
\be%
\lb{asbehan}
\ba{lll} %
\theta(p)=1.2992p-0.3364p^3+\cdots%
&\mbox{for } |p| \ll1 &\\[-0.15cm]%
&&\mbox{and}\\[-0.6cm]\\%
\theta(p)=\frac{\ds\pi}{\ds2}\,\mbox{sign}(p)-%
\frac{\ds\pi}{\ds\sqrt{6}}\frac{\ds1}{\ds{p^3}}-%
5.9502\frac{\ds1}{\ds{p^5}}+\cdots%
&\mbox{for } |p| \gg1\,. &%
\ea%
\ee%

\section{Conclusions}
The essential advantage of the suggested technique for
calculations of the $\theta (p)$ is that they are based on the
exact asymptotic behavior of the physical Bogolubov's angle
$\theta (p)$ at $p\rightarrow0$ and at $p\rightarrow \pm\infty$.
This gives us an opportunity to call the obtained solution for
$\theta (p)$ a semi-analytical one.

The accuracy of my calculations is $0.035$ or better. This error
appears when momentum $p$ is about either $0.4$ or $1.7$ momentum
units. However, when the absolute value of momentum is less than
$0.02$ or greater than $4.9$ momentum units, the accuracy has
already been an order of magnitude better, i.e. an accuracy about
$0.003$.

If less of the solution discrepancy of the integral equation for
$\theta(p)$ is needed, we suppose the accuracy of calculations can
be improved by adding terms of high order of $p$ to the trial
function of $\theta (p)$.

\end{document}